\documentclass[12pt]{article} 
\usepackage{a4} 
\usepackage{amssymb} 
\begin{document}

\newcommand{\Section}[1]{\setcounter{equation}{0}\section{#1}} 
\renewcommand{\theequation}{\thesection.\arabic{equation}} 
\def\be{\begin{equation}} 
\def\ee{\end{equation}} 
\def\bea{\begin{eqnarray}} 
\def\eea{\end{eqnarray}} 
\def\A{{\cal A}} 
\def\ve{\varepsilon} 
\def\ha{\frac{1}{2}} 
\def\hR{\hat R} 
\def\bx{\bar x} 
\def\by{\bar y} 
\def\l{\lambda} 
\def\hh{{1 \over 2}} 
\def\hq{\frac{1}{q}} 
\def\nn{\nonumber\\} 
\def\pat{\partial} 
\def\Ax{{\mathcal{A}_x}} 
\def\Az{{\mathcal{A}_z}} 
\def\dc{\stackrel{\diamond}{,}} 
\def\ds{\stackrel{*}{,}} 
\def\x{\hat x} 
\def\y{\hat y} 
\def\z{\hat z} 
\def\xx{{\hat X}} 
\def\aa{{\hat A}} 
\def\la{\lambda} 
\def\ka{\kappa}
\def\c#1{{\cal #1}} 
\def\ad{\mbox{ad}\,} 
 
  \newcommand*\ti[5]{{\em #5}, {#1} {\bf #2}, #3 (#4)} 
 
\providecommand{\href}[2]{#2} 
\newcommand*\xxx[1]{\href{http://xxx.lanl.gov/abs/#1}{{#1}}} 
 
\newcommand*\jhep{JHEP} 
\newcommand*\np{Nucl. Phys.} 
\newcommand*\pl{Phys. Lett.} 
\newcommand*\prl{Phys. Rev. Lett.} 
\newcommand*\pr{Phys. Rev.} 
\newcommand*\jmp{J. Math. Phys.} 

\title{$*$-Products on Quantum Spaces} 
 
\author{Hartmut Wachter\footnote{email: Hartmut.Wachter@Physik.Uni-Muenchen.de} 
and Michael Wohlgenannt\footnote{email: 
Michael.Wohlgenannt@Physik.Uni-Muenchen.de}\\ 
{\it Sektion Physik der Ludwig-Maximilians-Universit\"at}\\ 
{\it Theresienstr. 37, D-80333 M\"unchen}\\ 
} 
 
\date{} 
 
\maketitle 
 
\abstract{ 
In this paper we present explicit formulas for the $*$-product on quantum spaces 
which are of particular importance in physics, i.e., the $q$-deformed Minkowski 
space and the 
$q$-deformed Euclidean space in $3$ and $4$ dimensions, respectively. Our 
formulas are complete and formulated using the deformation parameter $q$. In 
addition, we worked out an expansion in powers of $h=\ln q$ up to second order, for  
all considered cases. 
}

\newpage

\section{Introduction}

Non-commutative space-time structures seem to be one of the most hopeful notions in
formulating finite quantum field theories \cite{grosse}. Even in the content of string
theory non-commutative geometries have recently been studied \cite{seiberg, berenstein}. 
Especially quantum
spaces which can lead to a lattice-like space-time structure provide a natural frame
work for a realistic non-commutative field theory \cite{lorek, ogievetsky}. In order to
do so we employ the $*$-product formalism which represents the non-commutative structure
on a commutative one \cite{madore,jurco, jurco2}.
\par
In the following we want to concern ourselves with coordinates which have quantum 
groups as their
underlying symmetry structure, in very much the same way as for example the
classical Minkowski space has the Lorentz group as its underlying symmetry
structure. Quantum groups are $q-$deformations of function algebras over classical
Lie groups (or $q-$deformations of the enveloping algebra of classical Lie 
algebras respectively) \cite{reshethikin}. The algebra generated 
by the coordinates is a comodule algebra of some quantum group and is called a quantum
space. So we can define the coordinate algebra $\mathcal{A}_q$ generated by
the coordinates $\hat{X}_1,\hat{X}_2,\dots,\hat{X}_n$ as 
\be
\mathcal{A}_q=\frac{\mathbb{C}<\hat{X}_1,\hat{X}_2,\dots,\hat{X}_n>}{\mathcal{R}},
\ee
where the relations between these coordinates reflect the quantum symmetry and therefore
determine the ideal $\mathcal{R}$. Formal power series in the coordinates are
allowed in $\mathcal{A}_q$. 
\par
The algebra $\mathcal{A}_q$ satisfies the Poincar$\acute{e}$-Birkhoff-Witt property,
i.e., the dimension of the subspace spanned by monomials of a fixed degree is
the same as the dimension of the subspace spanned by monomials in commutative
variables of the same degree. Taking this property and choosing the monomials of normal
ordering
$\hat{X}_1^{i_1}\hat{X}_2^{i_2}\dots\hat{X}_n^{i_n}$ as basis of
$\mathcal{A}_q$, we can establish an isomorphism between $\mathcal{A}_q$ and the
commutative algebra $\mathcal{A}$ generated by ordinary coordinates
$x_1,x_2,\dots,x_n$, as vector spaces.
\bea
W:\mathcal{A} &\longrightarrow & \mathcal{A}_q\\
W(x_1^{i_1}\dots x_n^{i_n}) & = & \hat{X}_1^{i_1}\dots\hat{X}_n^{i_n}.\nonumber
\eea
Let us consider a formal power series in the algebra $\mathcal{A}_q$,
$\hat{f}=\sum_ia_{i_1\dots i_n}\hat{X}_1^{i_1}\dots\hat{X}_n^{i_n}$,
the image under $W$ is $f=\sum_ia_{i_1\dots i_n}x_1^{i_1}\dots x_n^{i_n}$,
with the same coefficients $a_{i_1\dots i_n}$. This isomorphism of
vector spaces can be extended to an isomorphism of algebras introducing a
non-commutative product in $\mathcal{A}$, the socalled $*$-product. This product
is defined by the relation
\be
W(f*g)=W(f)W(g),
\ee
where $f$ and $g$ are formal power series in $\mathcal{A}$. This will give the
usual multiplication in the limit $q\to 1$ \cite{kontsevich, moyal, bayen}.
In order to do field theory on non-commutative spaces one needs to have a notion
of integration. This notion is given in \cite{steinacker}.
In this paper we will work out the necessary $*$-product for some relevant quantum
spaces, namely for the q-deformed $3$- and $4$-dimensional Euclidean space and
the q-deformed Minkowski space.

\section{q-Deformed $\mathbf{3}$-Dimensional Euclidean Space}

The algebra of the q-deformed version of $3$-dimensional Euclidean space is the 
algebra generated by the coordinates $\hat{X}_3$, $\hat{X}_+$, $\hat{X}_-$, 
satisfying the following relations \cite{lorek, cerchiai} 
\bea\label{relns} 
&\hat{X}_3\hat{X}_+=q^2\hat{X}_+\hat{X}_3,\quad 
\hat{X}_-\hat{X}_3  =  q^2\hat{X}_3\hat{X}_-,&\\ 
&\hat{X}_-\hat{X}_+=\hat{X}_+\hat{X}_-+\lambda\hat{X}_3\hat{X}_3,&\nonumber 
\eea  
where $\lambda=q-q^{-1}$. $q$ is the complex deformation parameter. 
\par 
As a basis of this quantum space we can take the monomials of normal ordering, 
$\hat{X}_+^{n_+}\hat{X}_3^{n_3}\hat{X}_-^{n_-}$. Using the isomorphism $W$  
introduced before, we can assign each 
monomial in commutative coordinates a normal ordered expression of non-commutative 
coordinates, 
\be 
W(x_+^{n_+}x_3^{n_3}x_-^{n_-})=\hat{X}_+^{n_+}\hat{X}_3^{n_3}\hat{X}_-^{n_-}. 
\ee 
The $*$-product on monomials is then defined by the condition 
\be\label{one} 
W\left((x_+^{n_+}x_3^{n_3}x_-^{n_-})*(x_+^{m_+}x_3^{m_3}x_-^{m_-})\right)  =  
W(x_+^{n_+}x_3^{n_3}x_-^{n_-})W(x_+^{m_+}x_3^{m_3}x_-^{m_-}) 
\ee 
The right hand side of (\ref{one}) has to be 
rewritten in normal ordering, using relations (\ref{relns}). For 
this aim, we need to calculate the commutation relations for 
$\hat{X}_-^{n_-}\hat{X}_3^{m_3}$, $\hat{X}_3^{n_3}\hat{X}_+^{m_+}$ and 
$\hat{X}_-^{n_-}\hat{X}_+^{m_+}$.\\  
These commutation relations read 
\bea\label{cr1} 
\hat{X}_-^{n_-}\hat{X}_3^{m_3} & = & q^{2n_-m_3}\hat{X}_3^{m_3}\hat{X}_-^{n_-},\\ 
\label{cr2} 
\hat{X}_3^{n_3}\hat{X}_+^{m_+} & = & 
q^{2n_3m_+}\hat{X}_+^{m_+}\hat{X}_3^{n_3},\\ 
\label{cr3} 
\hat{X}_-^{n_-}\hat{X}_+^{m_+} & = & \sum_{i=0}^{\mbox{{\tiny min}}\{n_-,m_+\}} 
\lambda^iB^{n_-,m_+}_i\hat{X}_+^{m_+-i}\hat{X}_3^{2i}\hat{X}_-^{n_--i}, 
\eea 
the coefficients $B_i^{n_-,m_+}$ satisfy the recursion relation 
\bea\label{four} 
B_0^{n_-,m_+} & = & 1,\nonumber\\ 
B_i^{n_-,m_+} & = & B_i^{n_-,m_+-1}+q^{4(m_+-i)}[[n_--(i-1)]]_{q^4} 
B_{i-1}^{n_-,m_+-1}, 
\eea 
where $[[n]]_{q^a}\equiv\frac{1-q^{an}}{1-q^a}$. As one can see by inserting  
(\ref{four}) has the solution  
\be\label{five} 
B_i^{n_-,m_+}=\frac{1}{[[i]]_{q^4}!}\frac{[[n_-]]_{q^4}!\,\, [[m_+]]_{q^4}!} 
{[[n_--i]]_{q^4}!\,\, [[m_+-i]]_{q^4}!}, 
\ee 
where $[[n]]_{q^a}!:= [[n]]_{q^a}[[n-1]]_{q^a}\cdot\dots\cdot[[1]]_{q^a}$, 
$[[0]]_{q^a}!:=1$.\\ 
(\ref{one}), (\ref{cr1}-\ref{cr3}) and (\ref{five}) together,  
yield the result 
\bea\label{six} 
W(x_+^{n_+}x_3^{n_3}x_-^{n_-}) && W(x_+^{m_+}x_3^{m_3}x_-^{m_-})=\nonumber\\ 
&=&\sum_{i=0}^{\mbox{{\tiny min}}\{m_+,n_-\}}C_i^{n_-,m_+}\quad 
        \hat{X}_+^{n_++m_+-i}\hat{X}_3^{n_3+m_3+2i}\hat{X}_-^{n_-+m_--i} 
        \nonumber\\ 
&=&W\left( \sum_{i=0}^{\mbox{{\tiny min}}\{m_+,n_-\}}C_i^{n_-,m_+}\quad 
        x_+^{n_++m_+-i}x_3^{n_3+m_3+2i}x_-^{n_-+m_--i}\right)\\ 
&=&W\left((x_+^{n_+}x_3^{n_3}x_-^{n_-})*(x_+^{m_+}x_3^{m_3}x_-^{m_-})\right), 
\nonumber 
\eea 
where $C_i^{n_-,m_+}=\lambda^iq^{2(n_3(m_+-i)+m_3(n_--i))}B_i^{n_-,m_+}$.\\ 
This is the $*$-product for monomials. In order to obtain the $*$-product for 
arbitrary formal power series 
$f=\sum_ia_{i_+,i_3,i_-}x_+^{i_+}x_3^{i_3}x_-^{i_-}$, we have to substitute 
\be\label{subst} 
q^{n_A}\quad \mbox{with} \quad 
q^{\hat{\sigma}_A}=q^{x_A\frac{\partial}{\partial x_A}},\quad A\in\{3,+,-\} 
\ee 
(no summation over $A$) with the usual commutative derivatives.\\ 
Applying this substitution to (\ref{six}) we end up at the expression 
\be\label{seven} 
f*g = 
\sum_{i=0}^\infty\lambda^i  \frac{x_3^{2i}}{[[i]]_{q^4}!}\,\,q^{2(\hat{\sigma}_3 
\hat{\sigma}'_++\hat{\sigma}_-\hat{\sigma}'_3)}\left(D_{q^4}^-\right)^i\!\! 
f(x)\cdot \left(D_{q^4}^+\right)^i\!\! g(x')\Big|_{x'\to x}, 
\ee 
$f,g\in\mathcal{A}_q$. 
We have used the $q$-differentiation operator 
$D_q^A\! f(x)=\frac{f(x_A)-f(qx_A)}{x_A-qx_A}$ (\cite{klimyk}) in the above formula. 
 
\par 
For practical purposes, we want to know an expansion of expressions (\ref{six}) 
and (\ref{seven}) in the variable $h=\ln q$. One expects that the main 
contribution to the $*$-product is made by the expansion coefficients up to $h^2$. 
So that we are not too far away from the classical situation, $h=0$.\\ 
For the expression (\ref{six}) we get the expansion 
\bea\label{exp1} 
(x_+^{n_+}x_3^{n_3}x_-^{n_-})&*&(x_+^{m_+}x_3^{m_3}x_-^{m_-})= 
  x_+^{n_++m_+}x_3^{n_3+m_3}x_-^{n_-+m_-}  \nonumber\\ 
& + & h \Big( a^{(1)}_0\!(\underline{n},\underline{m})\,\, x_+^{n_++m_+}x_3^{n_3+m_3} 
  x_-^{n_-+m_-} \nonumber\\ 
&& +\,\theta(n_-)\theta(m_+)\,\, a^{(1)}_1\!(\underline{n},\underline{m})\,\,  
  x_+^{n_++m_+-1}x_3^{n_3+m_3+2}x_-^{n_-+m_--1} \Big) \\ 
& + & h^2 \Big( a^{(2)}_0\! (\underline{n},\underline{m})\,\, 
  x_+^{n_++m_+}x_3^{n_3+m_3}x_-^{n_-+m_-}\nonumber\\ 
&& +\,\theta(n_-)\theta(m_+)\,\, a^{(2)}_1\! (\underline{n},\underline{m})\,\, 
  x_+^{n_++m_+-1}x_3^{n_3+m_3+2}x_-^{n_-+m_--1} \nonumber\\ 
&& +\, \theta(n_--1)\theta(m_+-1)\,\, a^{(2)}_2\! (\underline{n},\underline{m})\,\, 
  x_+^{n_++m_+-2}x_3^{n_3+m_3+4}x_-^{n_-+m_--2} \Big)\nonumber\\ 
& + & \mathcal{O}(h^3),\nonumber 
\eea 
where $\theta$ is the Heaviside function, and we have the coefficients 
\bea 
a^{(1)}_0(\underline{n},\underline{m})&=&2(n_3m_++m_3n_-),\nonumber\\ 
a^{(1)}_1(\underline{n},\underline{m})&=&2n_-m_+,\nonumber\\ 
a^{(2)}_0(\underline{n},\underline{m})&=&2(n_3m_++m_3n_-)^2,\\ 
a^{(2)}_1(\underline{n},\underline{m})&=&4n_-m_+((n_3+1)(m_+-1)+(m_3+1)(n_--1)), 
	\nonumber\\ 
a^{(2)}_2(\underline{n},\underline{m})&=&2n_-(n_--1)m_+(m_+-1).\nonumber 
\eea 
 
And in terms of derivatives we find 
\bea\label{equus1} 
f*g = && f(x)g(x)\nonumber\\ 
& + & h\Big(2(\hat\sigma_3\hat{\sigma}'_++\hat{\sigma}'_3\hat\sigma_-) 
  +2\frac{x_3^2}{x_+x_-}\hat\sigma_-\hat{\sigma}'_+ \Big)f(x)g(x')\Big|_{x'\to x} 
  \nonumber\\ 
& + & h^2\Big(2(\hat\sigma_3\hat{\sigma}'_++\hat{\sigma}'_3\hat\sigma_-)^2 
  +2\left(\frac{x_3^2}{x_+x_-}\right)^2 
  \hat\sigma_-(\hat\sigma_--1)\hat{\sigma}'_+(\hat{\sigma}'_+-1)\\ 
&& +4\frac{x_3^2}{x_+x_-} 
  \hat\sigma_-\hat{\sigma}'_+((\hat{\sigma}_3+1)(\hat{\sigma}'_+-1)+ 
  (\hat{\sigma}'_3+1)(\hat\sigma_--1))\Big) 
  f(x)g(x')\Big|_{x'\to x}\nonumber\\ 
& + & \mathcal{O}(h^3).\nonumber 
\eea

\section{q-Deformed $\mathbf{4}$-Dimensional Euclidean Space}

The procedure to get the $*$-product for the $4$-dimensional Euclidean space is
very much the same as in section 2. Therefore we will only state the results.
The quantum space algebra is freely generated by the coordinates $\hat{X}_1$,
$\hat{X}_2$, $\hat{X}_3$ and $\hat{X}_4$, divided by the ideal generated by the
following relations \cite{reshethikin, ocampo}
\bea
&\hat{X}_1\hat{X}_2=q\hat{X}_2\hat{X}_1,\,\,
\hat{X}_1\hat{X}_3=q\hat{X}_3\hat{X}_1,&\nonumber\\
&\hat{X}_3\hat{X}_4=q\hat{X}_4\hat{X}_3,\,\,
\hat{X}_2\hat{X}_4=q\hat{X}_4\hat{X}_2,&\\
&\hat{X}_2\hat{X}_3=\hat{X}_3\hat{X}_2,\,\,
\hat{X}_4\hat{X}_1-\hat{X}_1\hat{X}_4=\lambda\hat{X}_2\hat{X}_3.&\nonumber
\eea
As a basis we use the ordered monomials
$\hat{X}_1^{i_1}\hat{X}_2^{i_2}\hat{X}_3^{i_3}\hat{X}_4^{i_4}$, and
\be
W(x_1^{i_1}x_2^{i_2}x_3^{i_3}x_4^{i_4})=\hat{X}_1^{i_1}\hat{X}_2^{i_2}
\hat{X}_3^{i_3}\hat{X}_4^{i_4}.
\ee
We get to eqns. (\ref{cr1}), (\ref{cr2}) and (\ref{cr3}) analogue formulae 
\bea
&\hat{X}_2^{n_2}\hat{X}_1^{m_1}=q^{-n_2m_1}\hat{X}_1^{m_1}\hat{X}_2^{n_2},\,\,
\hat{X}_3^{n_3}\hat{X}_1^{m_1}=q^{-n_3m_1}\hat{X}_1^{m_1}\hat{X}_3^{n_3},&\nonumber\\
&\hat{X}_4^{n_4}\hat{X}_3^{m_3}=q^{-n_4m_3}\hat{X}_3^{m_3}\hat{X}_4^{n_4},\,\,
\hat{X}_4^{n_4}\hat{X}_2^{m_2}=q^{-n_4m_2}\hat{X}_2^{m_2}\hat{X}_4^{n_4},&\nonumber\\
&\hat{X}_3^{m_3}\hat{X}_2^{n_2}=\hat{X}_2^{n_2}\hat{X}_3^{m_3},&\\
&\hat{X}_4^{n_4}\hat{X}_1^{m_1}=\sum_{i=0}^{\mbox{\tiny{min}}\{n_4,m_1\}}\lambda^i
B_i^{n_4,m_1}\hat{X}_1^{m_1-i}\hat{X}_2^{i}\hat{X}_3^{i}\hat{X}_4^{n_4-i}&\nonumber
\eea
where 
\be
B_i^{n_4,m_1}=\frac{1}{[[i]]_{q^{-2}}!}\frac{[[n_4]]_{q^{-2}}!\,\, [[m_1]]_{q^{-2}}!}
{[[n_4-i]]_{q^{-2}}!\,\, [[m_1-i]]_{q^{-2}}!}.
\ee
Therefore the $*$-product of two monomials has the form
{\small 
\bea\label{4eucl} 
&&(x_1^{n_1}x_2^{n_2}x_3^{n_3}x_4^{n_4})*(x_1^{m_1}x_2^{m_2}x_3^{m_3}x_4^{m_4})=
\nonumber\\ 
&&=\sum_{i=0}^{\mbox{\tiny{min}}\{n_4,m_1\}}\lambda^i q^{-(n_2+n_3)(m_1-i)-
(m_2+m_3)(n_4-i)}B_i^{n_4,m_1}\times\\
&& \quad\times\,\, x_1^{n_1+m_1-i}x_2^{n_2+m_2+i}x_3^{n_3+m_3+i}
x_4^{n_4+m_4-i}.\nonumber
\eea}
Using again the substitution (\ref{subst}) we obtain for $f,g\in\mathcal{A}_q$
\be\label{4diff}
f*g=  
\sum_{i=0}^\infty \lambda^i \frac{\left( x_2x_3\right)^i}{[[i]]_{q^{-2}}!}\,\,
q^{-(\hat\sigma_2+\hat\sigma_3)\hat\sigma'_1- 
(\hat\sigma'_2+\hat\sigma'_3)\hat\sigma_4}\left( D^4_{q^{-2}}\right)^i\!\! f(x)\cdot 
\left( D^1_{q^{-2}}\right)^i\!\! g(x')\Big|_{x'\to x}, 
\ee 
with the same definitions and conventions as in the previous section. 
 
\par 
Again we want to expand expressions (\ref{4eucl}) and (\ref{4diff}) in terms of 
$h=\ln q$. We find  
\bea\label{find} 
(x_1^{n_1}x_2^{n_2}x_3^{n_3}x_4^{n_4})&*&(x_1^{m_1}x_2^{m_2}x_3^{m_3}x_4^{m_4})= 
   x_1^{n_1+m_1}x_2^{n_2+m_2}x_3^{n_3+m_3}x_4^{n_4+m_4}+\nonumber\\ 
& + & h\Big( 
   a_0^{(1)}\,\, x_1^{n_1+m_1}x_2^{n_2+m_2}x_3^{n_3+m_3}x_4^{n_4+m_4} 
   +\nonumber\\ 
&& +\,\theta(n_4)\theta(m_1)\,\, a_1^{(1)}\,\, 
   x_1^{n_1+m_1-1}x_2^{n_2+m_2+1}x_3^{n_3+m_3+1}x_4^{n_4+m_4-1}\Big)\\ 
& + & h^2\Big(a_0^{(2)}\,\, 
   x_1^{n_1+m_1}x_2^{n_2+m_2}x_3^{n_3+m_3}x_4^{n_4+m_4}\nonumber\\ 
&& +\,\theta(n_4)\theta(m_1)\,\, a_1^{(2)}\,\, 
   x_1^{n_1+m_1-1}x_2^{n_2+m_2+1}x_3^{n_3+m_3+1}x_4^{n_4+m_4-1}\nonumber\\ 
&& +\,\theta(n_4-1)\theta(m_1-1)\,\, a_2^{(2)}\,\, 
   x_1^{n_1+m_1-2}x_2^{n_2+m_2+2}x_3^{n_3+m_3+2}x_4^{n_4+m_4-2}\Big)\nonumber\\ 
& + & \mathcal{O}(h^3),\nonumber 
\eea 
where $a_i^{(j)}=a_i^{(j)}(\underline{n},\underline{m})$, 
\bea 
a^{(1)}_0(\underline{n},\underline{m})&=&-(n_2+n_3)m_1-(m_2+m_3)n_4
        ,\nonumber\\ 
a^{(1)}_1(\underline{n},\underline{m})&=&2n_4m_1,\nonumber\\ 
a^{(2)}_0(\underline{n},\underline{m})&=&{1\over 2}((n_2+n_3)m_1+(m_2+m_3)n_4)^2,\\ 
a^{(2)}_1(\underline{n},\underline{m})&=& 
   -2n_4m_1(((n_2+n_3)+1)(m_1-1)+((m_2+m_3)+1)(n_4-1)),\nonumber\\ 
a^{(2)}_2(\underline{n},\underline{m})&=&2n_4(n_4-1)m_1(m_1-1).\nonumber 
\eea 
And in terms of derivatives we find 
\bea 
f*g & = & f(x)g(x)\nonumber\\ 
& + & h\Big( 
  -(\hat\sigma_2+\hat\sigma_3)\hat{\sigma}'_1-(\hat{\sigma}'_2+\hat{\sigma}'_3) 
  \hat{\sigma}_4+2\frac{x_2x_3}{x_1x_4}\hat\sigma_4\hat{\sigma}'_1\Big) 
  f(x)g(x')\Big|_{x'\to x}\\ 
& + & h^2\Big({1\over 2}((\hat\sigma_2+\hat\sigma_3)\hat{\sigma}'_1+(\hat{\sigma}'_2+ 
  \hat{\sigma}'_3)\hat{\sigma}_4)^2+ 2\left(\frac{x_2x_3}{x_1x_4}\right)^2 
  \hat\sigma_4(\hat\sigma_4-1)\hat{\sigma'}_1(\hat{\sigma}'_1-1)\nonumber\\ 
&-&2\frac{x_2x_3}{x_1x_4}\hat\sigma_4\hat{\sigma}'_1
   \left(((\hat\sigma_2+\hat\sigma_3)+1)(\hat{\sigma}'_1-1)+ 
  ((\hat{\sigma}'_2+\hat{\sigma}'_3)+1)(\hat\sigma_4-1)\right) 
  \Big)f(x)g(x')\Big|_{x'\to x}\nonumber\\ 
& + & \mathcal{O}(h^3).\nonumber 
\eea 
The symmetry in all these expressions between $x_1$ and $x_4$, respectively $n_4$ 
and $m_1$ is remarkable. In eqn. (\ref{find}) the exponents of the variables $x_1$ 
and $x_4$ are always diminished by the same number. These powers are distributed 
symmetrically among the coordinates $x_2$ and $x_3$. 
This stems from the fact that $SO_q(4)$ can be decomposed 
into $2$ independent copies of $SU_q(2)$, as in the classical case.  
In case of the Lorentz group its decomposition leads also to the tensor product of  
$2$ copies of $SL_q(2)$, which are related to each 
other via complex conjugation. Thus we will not be able to observe this symmetry 
between the corresponding Minkowski coordinates, $x_0$ and $x_3$. Additional terms 
in (\ref{find}) will occur where the powers taken away from $x_-$ and $x_+$ are 
not symmetrically distributed among $x_0$ and $x_3$. But still some remnants of  
the symmetry are present, cf. (\ref{min10}).

\section{q-Deformed Minkowski Space}

The maybe most important case we want to discuss in this article is a  
$q$-deformed version of the Minkowski space, the co-module algebra of the  
$q$-deformed Lorentz group \cite{lorek, ogievetsky, lorek2, cerchiai2, carow}. 
$q$-Minkowski space is generated by the four coordinates $\hat{X}_0$, 
$\hat{X}_+$, $\hat{X}_3$ and $\hat{X}_-$, they satisfy the following relations 
\bea\label{min1} 
& \hat{X}_-\hat{X}_0=\hat{X}_0\hat{X}_-,\quad 
  \hat{X}_+\hat{X}_0=\hat{X}_0\hat{X}_+,\quad 
  \hat{X}_3\hat{X}_0=\hat{X}_0\hat{X}_3, & \nonumber\\ 
& \hat{X}_-\hat{X}_3-q^2\hat{X}_3\hat{X}_-=(1-q^2)\hat{X}_0\hat{X}_-,\quad 
  \hat{X}_3\hat{X}_+-q^2\hat{X}_+\hat{X}_3=(1-q^2)\hat{X}_0\hat{X}_+, & \\ 
& \hat{X}_-\hat{X}_+-\hat{X}_+\hat{X}_-=\lambda\left( 
\hat{X}_3\hat{X}_3-\hat{X}_0\hat{X}_3\right). & \nonumber 
\eea 
In order to make the calculations easier, we introduce a new set of coordinates 
$\hat{X}_0$, $\hat{X}_+$, $\hat{\widetilde{X}}_3$, $\hat{X}_-$, where 
\be\label{min2} 
\hat{\widetilde{X}}_3\equiv \hat{X}_3-\hat{X}_0. 
\ee 
Thus the relations (\ref{min1}) become 
\bea\label{min3} 
& \hat{X}_-\hat{\widetilde{X}}_3=q^2\hat{\widetilde{X}}_3\hat{X}_-,\quad 
  \hat{\widetilde{X}}_3\hat{X}_+=q^2\hat{X}_+\hat{\widetilde{X}}_3, & \\ 
& \hat{X}_-\hat{X}_+-\hat{X}_+\hat{X}_-=\lambda\left( 
\hat{\widetilde{X}}_3\hat{\widetilde{X}}_3+\hat{X}_0\hat{\widetilde{X}}_3 
\right). & \nonumber 
\eea 
We again introduce the isomorphism $W$ from the commutative coordinate algebra  
into the q-deformed Minkowski space  
\be 
W(x_0^{n_0}x_+^{n_+}\tilde{x}_3^{n_3}x_-^{n_-})=\hat{X}_0^{n_0}\hat{X}_+^{n_+} 
\hat{\widetilde{X}}_3^{n_3}\hat{X}_-^{n_-}, 
\ee 
the right hand side is defined as our normal ordering.\\ 
Using relations (\ref{min3}) we get 
\bea\label{min4} 
\hat{\widetilde{X}}_3^{n_3}\hat{X}_+^{m_+} & = & q^{2n_3m_+} 
   \hat{X}_+^{m_+}\hat{\widetilde{X}}_3^{n_3},\\ 
\hat{X}_-^{n_-}\hat{\widetilde{X}}_3^{m_3} & = & q^{2n_-m_3} 
   \hat{\widetilde{X}}_3^{m_3}\hat{X}_-^{n_-},\nonumber\\ 
\hat{X}_-^{n_-}\hat{X}_+^{m_+} & = & \sum_{i=0}^{\mbox{\tiny{min}}\{n_-,m_+\}} 
   \lambda^i \hat{X}_+^{m_+-i}F_i^{n_-,m_+}\! (\hat{X}_0,\hat{\widetilde{X}}_3)\, 
   \hat{X}_-^{n_--i},\nonumber    
\eea 
where the coefficients $F_i^{n,m}(\hat{X}_0,\hat{\widetilde{X}}_3)$ satisfy the 
recursion relation 
\bea\label{min5} 
F_i^{n,m}(\hat{X}_0,\hat{\widetilde{X}}_3) & = & 
  F_i^{n,m-1}(\hat{X}_0,\hat{\widetilde{X}}_3) 
  +F_{i-1}^{n,m-1}(\hat{X}_0,\hat{\widetilde{X}}_3)\times\nonumber\\ 
& \times & \left(q^{4(m-i)} 
  [[n-(i-1)]]_{q^4}\hat{\widetilde{X}}_3^2 +q^{2(m-i)}[[n-(i-1)]]_{q^2} 
  \hat{X}_0\hat{\widetilde{X}}_3\right),\nonumber\\ 
F_0^{n,m}(\hat{X}_0,\hat{\widetilde{X}}_3)& = & 1. 
\eea 
We could not deduce a closed expression for 
$F_i^{n,m}(\hat{X}_0,\hat{\widetilde{X}}_3)$ solving the recursion relations 
(\ref{min5}).  
\par 
However, we can write down what we have so far for the $*$-product  
of ordered monomials, 
\bea\label{min8} 
(x_0^{n_0}x_+^{n_+}\tilde{x}_3^{n_3}x_-^{n_-})&*& 
 (x_0^{m_0}x_+^{m_+}\tilde{x}_3^{m_3}x_-^{m_-})=\nonumber\\ 
&=&\sum_{i=0}^{\mbox{\tiny{min}}\{n_-,m_+\}}\lambda^iq^{2(n_3(m_+-i)+ 
 m_3(n_--i))}\times\\ 
&&\times\quad F_i^{n_-,m_+}\! (x_0,\tilde{x}_3)\, x_0^{n_0+m_0}x_+^{n_++m_+-i} 
 \tilde{x}_3^{n_3+m_3}x_-^{n_-+m_--i}.\nonumber 
\eea 
We can rewrite the recursion formula for $F_i^{n_-,m_+}(x_0,\tilde{x}_3)$ 
\bea\label{min6} 
F_j^{n,m} & = & \sum_{i=0}^{m-j}\left(q^{4i}[[n-(j-1)]]_{q^4} 
\tilde{x}_3^2+q^{2i}[[n-(j-1)]]_{q^2}x_0\tilde{x}_3 
\right)F_{j-1}^{n,i+(j-1)}\nonumber\\ 
\label{min7} 
& = & \sum_{i_0=0}^{m-j}\sum_{i_1=0}^{i_0}\cdots\sum_{i_{j-1}=0}^{i_{j-2}} 
\prod_{k=0}^{j-1}\sum_{l=0}^{n-(j-k)}\left(q^{4(l+i_k)}\tilde{x}_3^2 
+q^{2(l+i_k)}x_0\tilde{x}_3\right) 
\eea 
and expand this expression in powers of $h=\ln q$. 
The expansion of $F_i^{n_-,m_+}$ enables us to write down the $*$-product up  
to order $h^2$. In  
order to deduce a closed expression we will use the identification of the  
generators of $q$-deformed Minkowski space with combinations of the generators  
of the Drinfeld-Jimbo algebra $\mathcal{U}_q(sl_2)$ \cite{kulish, drinfeld, klimyk}. 
\par 
Expanding expression (\ref{min8}) in powers of $h$ reads 
\bea\label{min10} 
(x_0^{n_0}x_+^{n_+}\tilde{x}_3^{n_3}x_-^{n_-}) & * & 
  (x_0^{m_0}x_+^{m_+}\tilde{x}_3^{m_3}x_-^{m_-})= 
  x_0^{n_0+m_0}x_+^{n_++m_+}\tilde{x}_3^{n_3+m_3} 
  x_-^{n_-+m_-}\nonumber\\ 
& + & h\Big( a^{(1)}_{0,0}\! (\underline{n},\underline{m})\,\, x_0^{n_0+m_0} 
  x_+^{n_++m_+}\tilde{x}_3^{n_3+m_3}x_-^{n_-+m_-}\\ 
&&+ \,\theta(n_-)\theta(m_+)\sum_{i=0,1}a^{(1)}_{1-i,1+i}(\underline{n},\underline{m}) 
  \times\nonumber\\ 
&&\quad\times\quad x_0^{n_0+m_0+(1-i)}x_+^{n_++m_+-1}\tilde{x}_3^{n_3+m_3+(1+i)} 
  x_-^{n_-+m_--1}\Big)\nonumber\\ 
& + & h^2\Big(a^{(2)}_{0,0}\! (\underline{n},\underline{m})\,\, x_0^{n_0+m_0} 
  x_+^{n_++m_+}\tilde{x}_3^{n_3+m_3}x_-^{n_-+m_-}\nonumber\\ 
&& + \,\theta(n_-)\theta(m_+)\sum_{i=0,1}a^{(2)}_{1-i,1+i}(\underline{n}, 
  \underline{m})\times\nonumber\\ 
&&\quad\times\quad x_0^{n_0+m_0+(1-i)}x_+^{n_++m_+-1}\tilde{x}_3^{n_3+m_3+(1+i)} 
  x_-^{n_-+m_--1}\nonumber\\ 
&&+ \,\theta(n_--1)\theta(m_+-1)\sum_{i=0,1}a^{(2)}_{2-i,2+i}(\underline{n}, 
  \underline{m})\times\nonumber\\ 
&&\quad\times\quad x_0^{n_0+m_0+(2-i)}x_+^{n_++m_+-2}\tilde{x}_3^{n_3+m_3+(2+i)} 
  x_-^{n_-+m_--2}\Big)\nonumber\\ 
& + & \mathcal{O}(h^3),\nonumber 
\eea 
where 
\bea\label{min11} 
a^{(1)}_{0,0}(\underline{n},\underline{m})&=&2(n_3m_++m_3n_-),\nonumber\\ 
a^{(1)}_{1,1}(\underline{n},\underline{m})&=& 
   a^{(1)}_{0,2}(\underline{n},\underline{m})=2n_-m_+,\nonumber\\ 
a^{(2)}_{0,0}(\underline{n},\underline{m})&=&2(n_3m_++m_3n_-)^2,\nonumber\\ 
a^{(2)}_{1,1}(\underline{n},\underline{m})&=&2n_-m_+((2n_3+1)(m_+-1)+(2m_3+1) 
  (n_--1)),\\ 
a^{(2)}_{0,2}(\underline{n},\underline{m})&=&4n_-m_+((n_3+1)(m_+-1)+(m_3+1) 
  (n_--1)),\nonumber\\ 
a^{(2)}_{2,2}(\underline{n},\underline{m})&=&\frac{1}{2} 
  a^{(2)}_{1,3}(\underline{n},\underline{m})=a^{(2)}_{0,4}(\underline{n}, 
  \underline{m}) 
  =2n_-(n_--1)m_+(m_+-1).\nonumber 
\eea 
And in terms of derivatives we find 
\bea\label{min21} 
f*g= & & f(x)g(x)\nonumber\\ 
& + & h\Big( 2(\hat\sigma_3\hat\sigma'_++\hat\sigma'_3\hat\sigma_-)+ 
  2\frac{\tilde{x}_3^2+x_0\tilde{x}_3}{x_+x_-}\hat\sigma_-\hat\sigma'_+\Big) 
  f(x)g(x')\Big|_{x'\to x}\nonumber\\ 
& + & h^2\Big( 2(\hat\sigma_3\hat\sigma'_++\hat\sigma'_3\hat\sigma_-)^2+ 
4\frac{\tilde{x}_3^2}{x_+x_-}\hat\sigma_-\hat\sigma'_+((\hat\sigma_3+1) 
  (\hat\sigma'_+-1)+(\hat\sigma'_3+1)(\hat\sigma_--1))\nonumber\\ 
&& +\,2\frac{x_0\tilde{x}_3}{x_+x_-}\hat\sigma_-\hat\sigma'_+((2\hat\sigma_3+1) 
  (\hat\sigma'_+-1)+(2\hat\sigma'_3+1)(\hat\sigma_--1))\\ 
&& +\,2\left(\frac{\tilde{x}_3^2+x_0\tilde{x}_3}{x_+x_-}\right)^2 
  \hat\sigma_-(\hat\sigma_--1)\hat\sigma'_+(\hat\sigma'_+-1)\Big) 
  f(x)g(x')\Big|_{x'\to x}\nonumber\\ 
& + & \mathcal{O}(h^3).\nonumber
\eea

Finally, we want to deduce a closed
expression for the $*$-product (\ref{min8}). To this aim we have a look at the
algebra $\mathcal{U}_q(sl_2)$ \cite{klimyk}. The algebra is generated by the
four generators $E$, $F$, $K$, $K^{-1}$, satisfying the relations
\bea\label{min12}
&KE=q^2EK,\quad KF=q^{-2}FK,\quad KK^{-1}=K^{-1}K=1&,\nonumber\\
&EF-FE=\frac{K-K^{-1}}{q-q^{-1}}.
\eea
Further we have \cite{klimyk}
\bea\label{min15}
F^nE^m & = & E^mF^n\\
&&\!\! +\!\!\sum_{i=1}^{\mbox{\tiny{min}}\{n,m\}}(-\lambda)^{-i}
\frac{[n]!\,\, [m]!}{[i]!\,\, [n-i]!\,\, [m-i]!}\left( \prod_{j=0}^{i-1}
Kq^{n-m+j}-K^{-1}q^{-n+m-j}\right)\,E^{m-i}F^{n-i},\nonumber
\eea
where $[a]=\frac{q^a-q^{-a}}{q-q^{-1}}$.\\
The Operators $L_A$, $W$ defined in eqn. (\ref{min13})) can be 
interpreted as $q-$angular momentum operators \cite{lorek}. They span a proper 
subalgebra of $\mathcal{U}_q(su_2)$.
\bea\label{min13}
L_+ & \equiv & q^{-3}[2]^{-1/2}E,\nonumber\\
L_- & \equiv & -q^{-2}[2]^{-1/2}KF,\nonumber\\
L_3 & \equiv & q^{-3}[2]^{-1}(qFE-q^{-1}EF),\\
W & \equiv & K+q^3\lambda L_3.\nonumber
\eea
Because of (\ref{min12}), these generators satisfy the following relations
\bea\label{min14}
L_3L_+-q^2L_+L_3&=&-\frac{W}{q^2}L_+,\nonumber\\
L_-L_3-q^2L_3L_-&=&-\frac{W}{q^2}L_-,\nonumber\\
L_-L_+-L_+L_-&=&-\frac{W}{q^3}L_3+\lambda L_3L_3,\\
1=W^2-q^6\lambda^2(L_3L_3 & - & qL_+L_--q^{-1}L_-L_+).\nonumber
\eea
With the substitution $W\to q^3\lambda\hat{X}_0$, $L_A\to\hat{X}_A$,
$A\in\{+,3,-\}$, $\mathbf{1}\to q^6\lambda^2\hat{r}^2$ 
we regain the relations of $q$-Minkowski coordinates (\ref{min1}) 
\cite{blohmann}. 
Now we return to the third equation of (\ref{min4}). Using eqn. (\ref{min15}) 
one gets 
\bea\label{min16} 
\hat{X}_-^n\hat{X}_+^m=q^{2nm}\hat{X}_+^m\hat{X}_-^n & + & 
\sum_{i=1}^{\mbox{\tiny{min}}\{n,m\}}\frac{[n]!\,\, [m]!}{[i]!\,\, [n-i]!\,\, [m-i]!} 
\left(\frac{\lambda_-}{\lambda_+}\right)^i q^{2nm+i^2-2im}\\ 
& \times & \left(\prod_{k=0}^{i-1}q^{n-m+k}\hat{\widetilde{X}}_3^2-q^{-n+m-k} 
\hat{r}^2\right)\hat{X}_+^{m-i}\hat{X}_-^{n-i},\nonumber 
\eea 
where 
$\hat{r}^2=-q^{-2}\hat{\widetilde{X}}_3^2-(1+q^{-2})\hat{X}_0 
\hat{\widetilde{X}}_3+(q+q^{-1})\hat{X}_+\hat{X}_-$, and 
$\lambda_\pm=q\pm q^{-1}.$ The right hand side of eqn. 
(\ref{min16}) still has to be ordered according to the normal ordering. Note  
that $\hat{\widetilde{X}}_3^2$ and $\hat{r}^2$ commute, therefore we find 
\bea\label{min17} 
&&{\scriptstyle q^{i^2-2im}\prod_{k=0}^{i-1}\left(q^{n-m+k}\hat{\widetilde{X}}_3^2-q^{-n+m-k} 
  \,\hat{r}^2\right)\hat{X}_+^{m-i}\hat{X}_-^{n-i}=}\nonumber\\ 
&&{\scriptstyle =q^{-i^2}\sum_{k=0}^{i}(-1)^kq^{(1/2i-k)(i-1)}\left[\matrix{i\cr k}\right]_q 
  \left( 
  q^{i-2k}\hat{X}_+\right)^{m-i}\,\hat{\widetilde{X}}_3^{2(i-k)}\hat{r}^{2k}\, 
  \left(q^{i-2k}\hat{X}_-\right)^{n-i},} 
\eea 
where $\left[\matrix{i\cr k}\right]_q =\frac{[i]!}{[k]!\,\,[k-i]!}$.
One can also calculate 
$W^{-1}(\hat{\widetilde{X}}_3^{2(i-k)}\hat{r}^{2k})$ the last 
missing link to write down the $*$-product for q-deformed Minkowski space, and 
after a lengthy calculation one gets 
\be\label{min18} 
W^{-1}(\hat{\widetilde{X}}_3^{2(i-k)}\hat{r}^{2k})= 
\tilde{x}_3^{2(i-k)}\sum_{p=0}^k (q^{4(i-k)}\,\lambda_+x_+x_-)^p S_{k,p}(x_0,\tilde{x}_3), 
\ee 
where 
\bea\label{min19}  
S_{k,p}(x_0,\tilde{x}_3)& = & \left\{ \begin{array}{cc} 
 1, & \mbox{if } p=k \\ \sum\limits_{j_1=0}^p\sum\limits_{j_2=0}^{j_1}\cdots 
\sum\limits_{j_{k-p}=0}^{j_{k-p-1}}\prod_{l=1}^{k-p}a(x_0,q^{2j_l}\,\tilde{x}_3), 
 & \mbox{if } 0\le p<k 
 \end{array}\right.,\nonumber\\ 
&&\nonumber\\ 
a(x_0,\tilde{x}_3) & = & -q^{-2}\tilde{x}_3^2-(1+q^{-2})x_0\tilde{x}_3. 
\eea 
Eqns. (\ref{min16}), (\ref{min18}) and (\ref{min19}) enable us to order any two 
monomials in the $q$-Minkowski generators and to  
write down the $*$-product for q-deformed Minkowski space in a closed  
expression,
\bea\label{min20}
f*g & = & \sum_{i=0}^\infty \left( \frac{\lambda_-}{\lambda_+} \right)^i \sum_{k+j=i}  
 \frac{R_{k,j}\,(\underline{x})}{[[k]]_{q^2}!\,\, [[j]]_{q^2}!}\,\, 
 q^{(2\hat\sigma_3+\hat\sigma_-+i)\hat\sigma'_+ 
 +(2\hat\sigma'_3+\hat\sigma'_++i)\hat\sigma_-}\times\\ 
&\times&\left[ (D^-_{q^2})^i\! f\right] \!\! (x_0,x_+,\tilde{x}_3,q^{j-k}x_-)\cdot 
 \left[ (D^+_{q^2})^i\! g\right]\!\! (x'_0,q^{j-k}x'_+,\tilde{x}'_3,x'_-)\Big|_{x'\to x}, 
\nonumber
\eea
where $\underline{x}=(x_0,x_+,x_3,x_-)$ and with the polynomials
\bea
R_{k,j}(x_0,x_+,\tilde{x}_3,x_-) & = & (-q)^k
(q^j\tilde{x}_3^2)^j
\sum_{p=0}^kS_{k,p}(x_0,\tilde{x}_3)\,\lambda_+^p(q^{4j}x_+x_-)^p=\nonumber\\
& =& W^{-1}\!\left( (q^j\hat{\tilde{X_3}}^2)^j(-q\, \hat{r}^2)^k \right).
\eea
So finally, we have found both, the expansion of the $*$-product in powers of $h$ 
(\ref{min21}) and a closed expression (\ref{min20}).

\section{Remarks} 
 
Let us end with a few comments on eqns. (\ref{seven}), (\ref{4diff}) and 
(\ref{min20}). First of all, we can see
that the formulas for the $*$-product  have a similiar structure in all 
three cases. The commutative product is modified by an infinite sum of 
corrections,
\be
f*g=fg +\sum_{i=1}^\infty h^iB_i(f,g), 
\ee 
cf. \cite{bayen}. The $i^{\mbox{\tiny{th}}}$ term is of order  
$\mathcal{O}(\lambda^i)=\mathcal{O}(h^i)$. 
\par

Additionally, there are some kind of mixed scaling operators of the form 
$q^{a\hat{\sigma}'\hat\sigma}$, which lead to a displacement effect. The
derivatives in the exponent will shift the argument of the function, such that
the value of the $*$-product at a given point depends not only on their values 
at that single point.
The displacement effect is present in all dimensions and shows that
non-commutativity induced by $q-$deformation implies some kind of non-locality.
Especially in Minkowski space, one is forced to reinterpret the concept of
causality, as the $*-$product, which can be considered as some kind of
interaction, does not only depend on the nearby past but also on the nearby
future.
\par

The remaining operators and factors are responsible for an effect we have
already mentioned at the end of section $3$. This substitution effect is absent
in less than $3$ dimensions. It transforms the (plane) coordinates $X_+$ and
$X_-$ ($X_1$ and $X_4$ resp.) into the transverse coordinate $X_3$ and the time
coordinate $X_0$ ($X_2$ and $X_3$ resp.). It also shows that physical quantities
like charge densities initially restricted to a plane may expand in transverse
directions or undergo a mysterious evolution in time.

\subsubsection*{Acknoledgement}

First of all we want to express our gratitude to Julius Wess for his efforts, 
suggestions and discussions. And we would like to thank Fabian Bachmaier, Peter 
Schupp and Christian Blohmann for useful discussions and their steady support.


\begin{thebibliography}{30}

\bibitem{grosse}
H. Grosse, C. Klim$\check{c}$ik, P. Pre$\check{s}$najder, {\it Towards finite 
quantum field theory
in non-commutative geometry}, Int.J.Theor.Phys. {\bf 35} (1996) 231, 
hep-th/9505175.

\bibitem{seiberg}
N. Seiberg, E. Witten, {\it String Theory on Noncommutative Geometry}, JHEP
{\bf 9909}, 032 (1999), hep-th/$9908142$.

\bibitem{berenstein}
D. Berenstein, V. Jejjala, R. Leigh, {\it Marginal and Relevant Deformations of N=4 Field
Theories and Non-Commutative Moduli Spaces of Vacua}, Nucl.Phys. B{\bf 589}  (2000) 196,
hep-th/0005087.\\
{\it Non-Commutative Moduli Spaces, Dielectric Tori and T-Duality}, Phys.Lett. B{\bf 493}
(2000) 162, hep-th/0006168.


\bibitem{lorek}
A. Lorek, W. Weich, J. Wess, {\it Non Commutative Euclidean and Minkowski
Structures}, Z.Phys. C{\bf 76} (1997) 375, q-alg/9702025.

\bibitem{ogievetsky}
O. Ogievetsky, W. B. Schmittke, J. Wess, B. Zumino, {\it q-Deformed 
Poincar$\acute{e}$ Algebra}, Commun.Math.Phys. {\bf 150} (1992) 495.

\bibitem{madore}
J. Madore, S. Schraml, P. Schupp, J. Wess, {\it Gauge Theory on Noncommutative
Spaces}, Eur.Phys.J. C{\bf 16} (2000) 161,
    hep-th/$0001203$.

\bibitem{jurco}
B. Jur$\check{c}$o, P. Schupp, {\it Non-commutative Yang-Mills theory from equivalence of
star products}, Eur.Phys.J. C{\bf 14} (2000) 367, hep-th/0001032.

\bibitem{jurco2}
B. Jur$\check{c}$o, P. Schupp, J. Wess, {\it Non-commutative gauge theory for Poisson
manifolds}, Nucl.Phys. B{\bf 584} (2000) 784, hep-th/0005005.

\bibitem{reshethikin}
N.Yu. Reshetikhin, L.A. Takhtadzhyan, L.D. Faddeev, {\it Quantization of Lie
Groups and Lie Algebras}, Leningrad Math.J. {\bf 1} (1990) 193.

\bibitem{kontsevich}
M. Kontsevich, {\it Deformation Quantization of Poisson Manifolds, I},
q-alg/9709040.

\bibitem{moyal}
J.E. Moyal, {\it Quantum mechanics as a statistical theory}, Proc.Camb.Phil.Soc. {\bf
45} (1949) 99.

\bibitem{steinacker}
Harold Steinacker, {\it Integration on quantum Euclidean space and sphere in N
dimensions}, q-alg/9506020.

\bibitem{cerchiai}
B. L. Cerchiai, J. Madore, S. Schraml, J. Wess, {\it Structure of the
Three-dimensional Quantum Euclidean Space}, Eur.Phys.J. C{\bf 16} (2000) 169,
math.QA/0004011.

\bibitem{klimyk}
A. Klimyk, K. Schm\"udgen, {\it Quantum Groups and their Representations}, chapter
2, 3 resp., Springer Verlag, Berlin (1997).

\bibitem{ocampo}
H. Ocampo, {\it $SO_q(4)$ quantum mechanics}, Z.Phys. C{\bf 70} (1996) 525.

\bibitem{lorek2}
A. Lorek, W. B. Schmittke, J. Wess, {\it $SU_q(2)$ Covariant
$\hat\mathcal{R}$-Matrices for Reducible Representations}, Lett.Math.Phys. {\bf
31} (1994) 279.

\bibitem{cerchiai2}
B. L. Cerchiai, J. Wess, {\it q-deformed Minkowski Space Based on a q-Lorentz
Algebra}, Eur.Phys.J. C{\bf 5} (1998) 553, math.QA/9801104.

\bibitem{carow}
U. Carow-Watamura, M. Schlieker, M. Scholl, S. Watamura, {\it Tensor
representation of the quantum group $SL_q(2,C)$ and quantum Minkowski space},
Z.Phys. C{\bf 48} (1990) 159.\\
{\it A Quantum Lorentz Group}, Int.Journ.Mod.Phys. A{\bf 6} (1991) 3081.
 
\bibitem{kulish}
P.P. Kulish, N.Yu. Reshetikhin, {\it Quantum linear problem for the Sine-Gordon equation
and higher representations}, Zap.Nauchn.Sem. LOMI {\bf 101} (1981) 101.

\bibitem{drinfeld}
V.G. Drinfeld, {\it Quantum groups}, Proc. of the Int. Congress of Math. (A.M. Gleason,
ed.), Amer.Math.Soc., Providence (1986), pp 798-826.

\bibitem{blohmann}
Chr. Blohmann, {\it Spin Representations of the q-Poincar$\acute{e}$
Algebra}, Ph.D. thesis, section 3.2.1, Ludwig-Maximilians-Universit\"at M\"unchen, Fakult\"at
f\"ur Physik (2001), math.QA/0110219.

\bibitem{bayen}
F. Bayen, M. Flato, C. Fr{\o}nsdal, A. Lichnerowicz, D. Sternheimer, {\it Deformation theory and quantization. I.
Deformations of symplectic structures}, Ann.Phys. {\bf 111} (1978), no. 1, 61.

\end{thebibliography}
\end{document}